\documentclass[11pt]{article}
\usepackage[utf8]{inputenc}
\pdfoutput=1
\usepackage{amssymb,amsmath,mathrsfs,enumerate}
\usepackage{graphicx,rotate,multicol}
\usepackage{float}
\usepackage{tocloft}
\usepackage{subcaption}
\usepackage[margin=10pt,labelfont=bf]{caption}
\usepackage{cite}
\usepackage{soul}
\usepackage[colorlinks=true,
            linkcolor=blue,
            urlcolor=blue,
            citecolor=teal]{hyperref}
\usepackage{multirow}
\usepackage{placeins}
\usepackage[normalem]{ulem}
\makeatletter
\let\Hy@linktoc\Hy@linktoc@page
\makeatother
\usepackage{color}
\definecolor{ourcolor}{rgb}{0.7, 0.25, 0.05}
\usepackage{tikz,braket}
\long\def\rpl#1!!#2!!{\textcolor{red}{#1} \textcolor{blue}{#2}}

\def \order(#1){{\mathcal O} \left(#1 \right)}

\evensidemargin=\oddsidemargin
\allowdisplaybreaks

\title{\color{black}{\bf Constraining dark matter self-interaction from kinetic heating in neutron stars}}

\author {\bf Sambo Sarkar$^{a,b}$\footnote{sambosarkar92@gmail.com}
	\\[10pt]
\small\em $^a$Department of Physical Sciences, Indian Institute of Science Education and Research Berhampur,\\
\small\em Ganjam, Odisha, 760003, India\\
\small\em $^b$Laboratory for Symmetry and Structure of the Universe, Department of Physics,\\
\small\em Jeonbuk National University, Jeonju 54896, Korea\\
}
\date{}
\begin{document} 
	
	\maketitle

	\begin{abstract}
		
		Dark matter search strategies have started advancing towards the neutrino fog. In this regard, compact objects such as neutron stars have already demonstrated their ability in probing such low DM-nucleon cross-sections from dark matter induced effects. In the optically thin limit, effect of dark matter self-interaction becomes relevant and may assist the capture and thermalization of dark matter inside stars, imparting observable changes on neutron star temperatures. The resulting radiation although weak can be potentially detected by the James Webb Space Telescope and upcoming Thirty Meter Telescope and the European Extremely Large Telescope. Observation of cold neutron stars accompanied by advancements in direct detection probes would provide stringent constraints  or a smoking-gun signature for dark matter self-interactions. The potential detection of a neutron star with surface temperatures $\sim (1000-1200)$ K in the optically thin limit can push the bounds on asymmetric dark matter self-interaction cross-section to approximately two orders of magnitude more stringent than the bullet cluster.
		    
	\end{abstract}

\section{Introduction}
\label{sec:intro}

The existence and predominance of dark matter (DM) has been validated time and again \cite{Clowe:2006eq,Planck:2018vyg}, with the possibility for DM to be particulate in nature and having tiny interactions with the standard model (SM) sector \cite{Goodman:1984dc,Bertone:2010zza,Khlopov:2013ava,Gross:2018ivp,Arcadi:2019lka}, and the dark sector \cite{delCampo:2015vha,Aoki:2016glu,SantanaJunior:2024cug,Kamada:2021muh}. The last few decades have witnessed a surge in studies concerning the detection and interaction of DM from terrestrial detectors \cite{Cui:2017nnn,XENON:2018voc}, and also using astrophysical objects \cite{Iocco:2008xb,Bramante:2021dyx}. The capture and subsequent annihilation of DM inside celestial bodies has proven to be a promising technique to detect or constrain its particle properties \cite{Slatyer:2017sev,Leane:2020liq,Boddy:2022knd,Ema:2024wqr}.

As astrophysical bodies move within a DM halo, the DM particles get attracted and subsequently trapped by the gravitational potential of the astrophysical objects. Depending on the mass of DM, single or multiple scattering events can reduce the incumbent energy of DM particles below the escape energy of the concerned celestial object \cite{Press1985CaptureBT,Gould:1987ir,Gould:1987ju,Bose:2022ola}. On its entry into the stellar interior, DM starts to scatter from the constituents of these celestial objects. Non-gravitational interactions with SM particles cause a sufficient reduction in DM energy. This process in literature is referred to as DM capture. The accumulated DM particles will thermalize at the core and may annihilate to SM or dark sector states, leading to interesting phenomenological signals from neutron stars (NS) \cite{Kouvaris:2007ay,Bramante:2017xlb,Bell:2018pkk,Acevedo:2019agu,Joglekar:2019vzy,Bell:2020jou,Joglekar:2020liw,Bell:2020lmm,Ilie:2020vec,Maity:2021fxw,Ilie:2021iyh,Anzuini:2021lnv,McKeen:2021jbh,Coffey:2022eav,Nguyen:2022zwb}, white dwarfs \cite{McCullough:2010ai,Dasgupta:2019juq,Bell:2021fye}, exoplanets \cite{Leane:2020wob,Benito:2024yki,Phoroutan-Mehr:2025hjz}, Earth \cite{Chauhan:2016joa,Bramante:2022pmn}, and the Sun \cite{Bell:2011sn,Busoni:2017mhe,Garani:2017jcj,Bell:2021esh,Maity:2023rez,Nguyen:2026nhe}. Final state SM products or long-lived dark mediators from DM annihilation can lead to an observable neutrino or gamma-ray flux at terrestrial detectors \cite{PhysRevLett.55.257,IceCube:2012ugg,Pospelov:2008jd,Batell:2009zp,Leane:2017vag,Bell:2021pyy,Leane:2021ihh,Krishna:2025ncv}. The accumulated DM particles inside a celestial object may lead to ignition of supernova \cite{Bramante:2015cua,Janish:2019nkk}, formation of black holes \cite{PhysRevD.40.3221,GOULD1990337,Bertone:2007ae,Bell:2013xk,Garani:2018kkd,Acevedo:2020gro,Dasgupta:2020mqg,Ray:2023auh}, and alterations in periods of binary pulsars \cite{Hassani:2020uhd}. Such processes however remain associated with astrophysical uncertainties \cite{Zhu:2019oax,Nunez-Castineyra:2019odi,Lopes:2020dau,Bose:2022ola,Chatterjee:2022dhp}. 

Neutron stars have provided some of the stringent constraints on DM-nucleon scattering cross-section, comparable to direct DM search \cite{Bertone:2004pz,Raj:2017wrv,Bose:2021yhz,Liu:2025qco}. Inside a NS, the DM accumulated through scattering with nucleons or leptons can further assist the capture and thermalization of DM in the presence of DM self-interactions \cite{Zentner:2009is}. DM self-interactions which was originally proposed to address thermalization driven core formation in galaxies \cite{Spergel:1999mh}, has been found to have interesting phenomenological consequences in compact objects \cite{Mariani:2023wtv} and main sequence stars \cite{Chen:2015uha}. Some of the competitive and early constraints on self-interacting DM (SIDM) have been presented using cosmological probes as gravitational lensing, X-ray spectroscopy and dwarf galaxies \cite{Sean:2017sea,Ebisu:2021bjh,Ray:2022ydr,Ray:2025xrv}. DM self-interactions in neutron stars therefore has been a topic of interest \cite{Bramante:2013nma,Lin:2020zmm,Dasgupta:2020dik}. In this work we consider self-interaction driven capture and consequent kinetic heating inside a NS by asymmetric DM \cite{Petraki:2013wwa,Dey:2025atz}. Thereby presenting bounds on the specific DM self-interaction cross-section for GeV scale masses. We show that neutron stars can be used to provide limits on the specific SIDM cross-section $\sigma/m$, orders of magnitude stringent than the bounds obtained from Bullet cluster \cite{Robertson:2016xjh}.

This paper is organized as follows. In section \ref{sec:formalism} we visit the numerical rationale for DM capture via nucleons and DM self-interactions. In section \ref{sec:numberevolve} we calculate the complete evolution of captured DM in presence of self-interactions. We discuss dark kinetic heating and its implication on the standard NS cooling scenario in section \ref{sec:Kineticheat}. In section \ref{sec:detec}, we discuss the prospects of probing dark kinetic heating in relatively cold neutron stars and provide constraints on $\sigma/m$. Finally we summarize and conclude in section \ref{sec:conclude}. 
\section{Dark matter capture in neutron stars}
\label{sec:formalism}

A recap of DM capture formalism in neutron stars is now in order. As NS orbits within the DM halo, its strong gravitational field attracts DM particles, leading to an enhanced DM density around the NS. Upon reaching the NS surface, DM particles gain a velocity $w = \sqrt{u_{\chi}^2+v_{\rm esc}^2}$, with $v_{\rm esc}$ being the NS escape velocity and $u_{\chi}$ the incumbent DM velocity. After few collisions with the NS constituents, DM enters the surface of the star ($r_{\chi}<R_*$), with their velocities becoming less than $v_{\rm esc}$. Here $r_{\chi}$ is the orbital radius of DM and $R_*$ radius of the NS. Once inside, DM scatters with neutrons and other constituent particles, thereby thermalizing and gradually migrating to the NS core. For DM masses less than neutron mass, effect of Pauli blocking needs to be taken into account \cite{Bertoni:2013bsa}. The capture rate is related to the DM phase space distribution around the NS in consideration, whose standard choice follows a Maxwell-Boltzmann (MB) profile with the matter density being scaled with radius \cite{Press1985CaptureBT,Gould:1987ir,Kouvaris:2007ay,Bramante:2017xlb}. Alternate empirical and non-thermal distributions can also be used to quantify the DM velocity distribution \cite{Bose:2022ola}.
\subsection{Capture by neutron scattering}
\label{subsec:nucleoncap}

Here we revisit DM capture through the scattering with neutrons inside the star. As we are concerned with DM masses less than TeV, we use the single scatter capture formalism with neutrons as the major target. The capture rate is given by \cite{PhysRevD.40.3221,Kouvaris:2007ay,Bell:2020jou},
\begin{equation}
\label{eq:caprate}
C_{\rm c} = \frac{4 \pi}{v_*} \frac{\rho_{\chi}}{m_{\chi}} \text{Erf}\left(\sqrt{1.5} \frac{v_*}{v_d} \right) \int_0 ^{R_{*}} \frac{\sqrt{1-B(r)}}{B(r)}\Omega^-(r) r^2 dr.
\end{equation}
Here $\rho_{\chi}$, $v_{\rm d}$ and $v_{*}$ are the ambient DM density, DM velocity dispersion, and the NS velocity respectively. We take them to be $0.4\,\rm GeV/cm^3$, $270 \rm \, km/s$ and $233 \rm \, km/s$ respectively. $m_{\chi}$ and $\Omega^-(r)$ being the DM mass and the inward interaction rate of DM  respectively. $B(r)$ is the time component of the Schwarzschild metric\footnote{In order to draw parallels between the relativistic approach and non-relativistic capture rates, one can replace the factor $1 - B(r)$ with the escape velocity $v_{\rm esc}^2$.} with $1/B(r)$ as a relativistic correction factor \cite{Bell:2020jou}. DM captured by leptons can also give rise to kinetic heating in NS, with loop induced coupling to nucleons \cite{Bell:2019pyc}
\subsection{Capture by self-scattering}
\label{subsec:selfcap}

The accumulated and thermalized DM inside the NS can further help to enhance the number of captured DM particles in presence of self-scattering. These captured DM now act as additional targets for energy deposition. The DM self-capture rate $C_s$ is given by \cite{Kouvaris:2007ay,Zentner:2009is,Guver:2012ba},
\begin{equation}
\label{eq:selfcaprate}
C_{\rm s} = \frac{v_{\rm esc}^2(R_{*})}{ v_*} \frac{\rho_{\chi}}{m_{\chi}} \text{Erf} \left( \sqrt{1.5}\frac{v_*}{v_d} \right) \left\langle \phi \right\rangle \frac{\sigma_{\chi \chi}}{B(R_{*})} ,
\end{equation}
where the symbols carry their usual meaning. Here $\sigma_{\chi \chi}$ is the DM self-scattering cross-section. The quantity $\left\langle \phi \right\rangle$ signifies compactness of the star \cite{Zentner:2009is}, which we take as unity\footnote{This comes from the assumption of matter density inside the NS being uniform. However, density being greater at the center makes this estimate conservative.}, a conservative approach following \cite{Guver:2012ba, Chen:2018ohx}.
\section{Accumulation of DM particles}
\label{sec:numberevolve}

\begin{figure}[t]
	\begin{center}
	\includegraphics[scale=0.35]{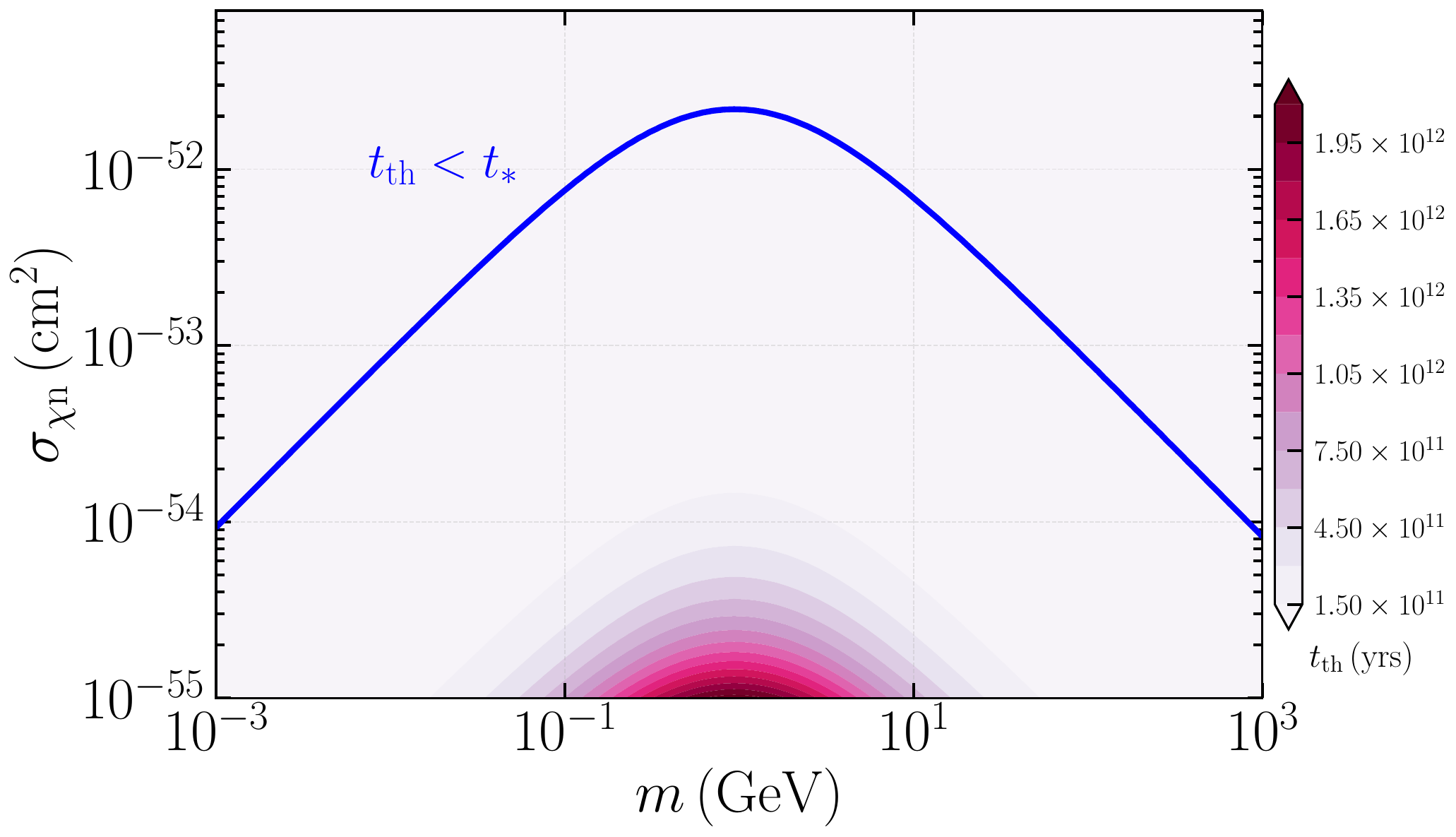}
	\end{center}
	\caption{Contour plot showing the region of $\sigma_{\chi \rm n}$ vs DM mass space, parameterized over the DM thermalization time. Region above the blue curve shows the parameter space for which thermalization timescales are shorter than the NS age of 1 Gyr.}
	\label{fig:thermscale}
\end{figure}

With the ingredients in place, we now evaluate the temporal variation of DM number $N_{\chi}(t)$, inside a NS. The DM undergoes numerous scattering with the neutrons gradually reducing its energy to temperatures of the NS core. Energies transferred to the neutrons heat up the star in process. In literature this is referred to as dark kinetic heating. Time required for DM to thermalize and effectively deposit its energy in the NS is defined as DM thermalization timescale $t_{\rm th}$. For the case of a constant DM-neutron cross-section, the thermalization time can be estimated using \cite{Bertoni:2013bsa,Bramante:2023djs,Bell:2023ysh}
\begin{equation}
\label{eq:ttherm}
t_{\rm th} \sim 3\times10^3 \frac{m_{\rm n}
	\,m_{\chi}}{(m_{\chi}+m_{\rm n})^2}\left( \frac{2\times10^{-45}\, \rm cm^2}{\sigma_{\chi \rm n}}\right) \left( \frac{10^5\,\rm  K}{T_{\rm NS}}\right)\, \rm yr,
\end{equation}
where the symbols carry their usual meaning, with $T_{\rm NS}$ as the effective NS core temperature.
In figure \ref{fig:thermscale}, we plot the contours for thermalization timescales with a typical core temperature of $10^5$ K and velocity-independent DM-nucleon scattering. Although in this work we consider the impact of DM self-capture in neutron stars, the expression for thermalization time remains independent of $\sigma_{\chi \chi}$. After thermalization the orbits of DM particles shrink from $\sim 10$ km, and settles into a dense core with radius $\sim 20$ cm, at the NS core \cite{McDermott:2011jp}. The volume of this DM core being extremely small compared to the NS, DM particles which are subsequently captured rely on the ubiquitous nucleons for thermalization. Therefore we consider equation \eqref{eq:ttherm} as the estimate for DM thermalization timescales even in the presence of DM self-interactions.  We use equation \eqref{eq:ttherm} to scan the DM mass $m_{\chi}$, vs. DM-nucleon scattering cross-section $\sigma_{\chi \rm n}$, parameter space, over the DM thermalization timescales. We see for cross-sections $\sim 10^{-54}\rm cm^2$ and $m_{\chi} \sim 100$ GeV, the thermalization timescales are larger than the typical age of a NS. 
As the DM accumulates inside the core, they can annihilate with itself or co-annihilate with SM particles. This acts as an additional source for NS heating by DM. Expectedly this effect is absent for asymmetric DM for which NS can attain a maximum temperatures of $1750$ K \cite{Baryakhtar:2017dbj}. In the absence of annihilation (i.e. $C_{\rm a}$=0), the evolution of asymmetric DM inside a typical NS is given by
\begin{equation}
\label{eq:Nevolve}
\frac{dN_{\chi}}{dt} =C_{\rm c}+C_{\rm s} N_{\chi},
\end{equation}
where the symbols carry their usual meaning. For our DM mass range of interest we have safely neglected DM evaporation. We solve equation \eqref{eq:Nevolve} in phases, depending on the saturation timescales $t_{\rm G}$, and the DM thermalization timescales $t_{\rm th}$, and determine the evolution of $N_{\chi}(t)$. The capture rate via neutrons is proportional to $\sigma_{\chi \rm n}$, which saturates at a value $\sigma_{\rm sat} \sim \pi R_*^2 m_{\rm n}/M_*$ \cite{Guver:2012ba}. Here $m_{\rm n}$ is the mass of a neutron and $M_*$ is the mass of NS.

In presence of DM self-interactions, the number of DM particles captured increases exponentially till a saturation point. This saturation timescale $t_{\rm G}$ for given $\sigma_{\chi \chi}$, is defined by $N_{\chi}(t_{\rm G}) \sim \pi r_{\chi}^2 (t_{\rm G})/ \sigma_{\chi \chi}$ \cite{Guver:2012ba}. Based on equation \eqref{eq:Nevolve}, evolution of $N_{\chi}$ till $t \leq t_{\rm G}$ is given by,
\begin{equation}
\label{eq:N1}
N_{\chi}(t)=\frac{C_{\rm c}}{C_{\rm s}}(e^{C_{\rm s} t}).
\end{equation}
For $t_{\rm {\rm G}}<t<t_{\rm th}$, the term involving self-capture in equation \eqref{eq:Nevolve} gets modified as $C_{\rm s} N_{\chi}(t_{\rm {\rm G}})\frac{t}{t_{\rm {\rm G}}}$. The solution to which is given by 
\begin{equation}
\label{eq:N2}
N_{\chi}(t)= (t-t_{\rm G})C_{\rm c}+ C_{\rm s} N_{\chi}(t_{\rm G}) t_{\rm G} {\rm ln} \frac{t}{t_{\rm G}}.
\end{equation}
After thermalization $t>t_{\rm th}$, the self-capture term in equation \eqref{eq:Nevolve} gets modified as $C_{\rm s} \pi r_{\rm th}^2/\sigma_{\chi \chi}$. The solution to which takes the form
\begin{equation}
\label{eq:N3}
N_{\chi}(t)= \frac{\pi r_{\chi}^2(t_{\rm G})}{\sigma_{\chi \chi}}.
\end{equation}
Therefore, a NS of age $t_{*}$, accumulates $N_{\chi}(t_{*})$ particles obtained by summing over equations \eqref{eq:N1}, \eqref{eq:N2} and \eqref{eq:N3}.
\begin{figure}[t!]
	\begin{center}
		\includegraphics[scale=0.25]{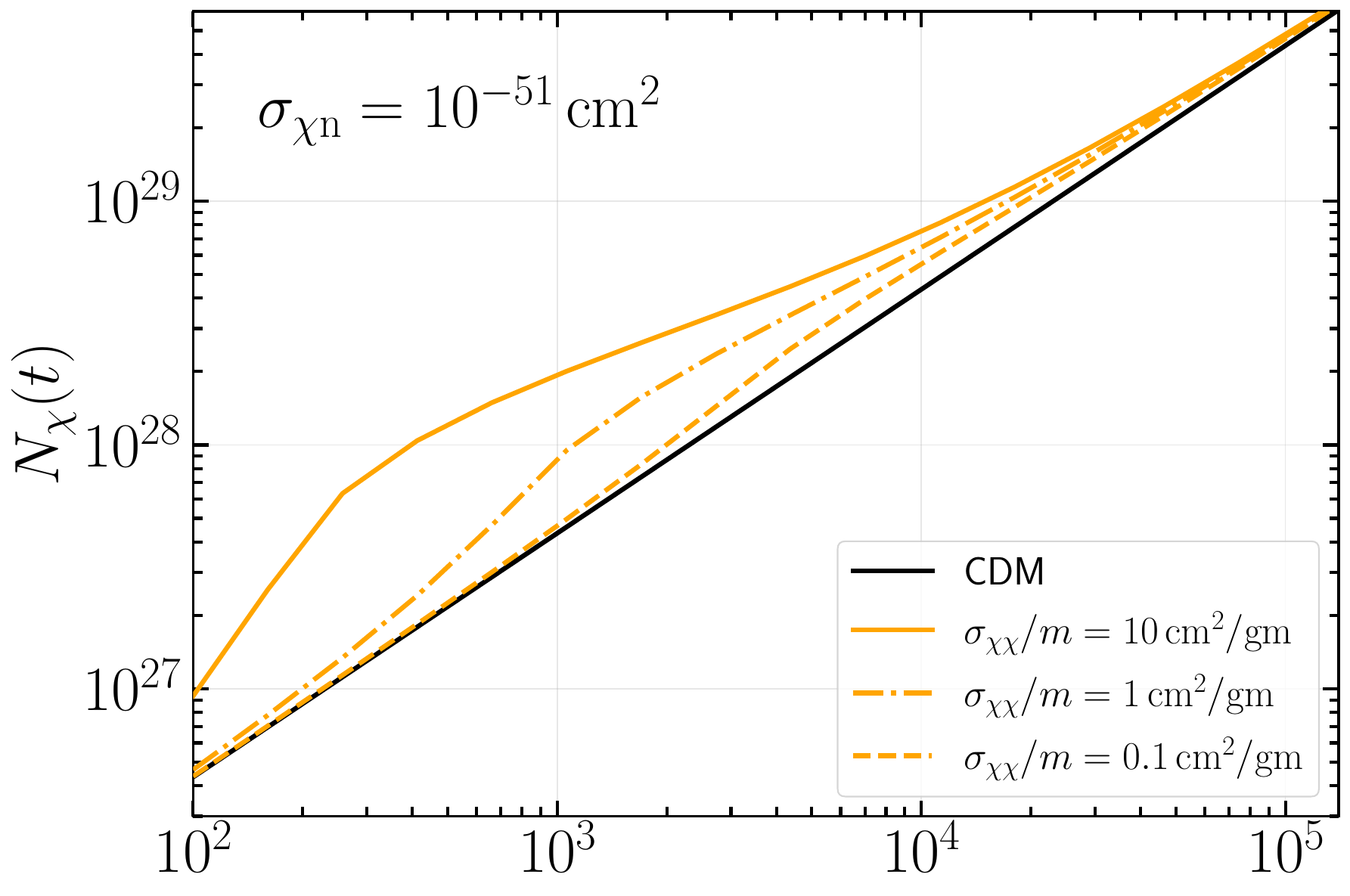}
		\includegraphics[scale=0.25]{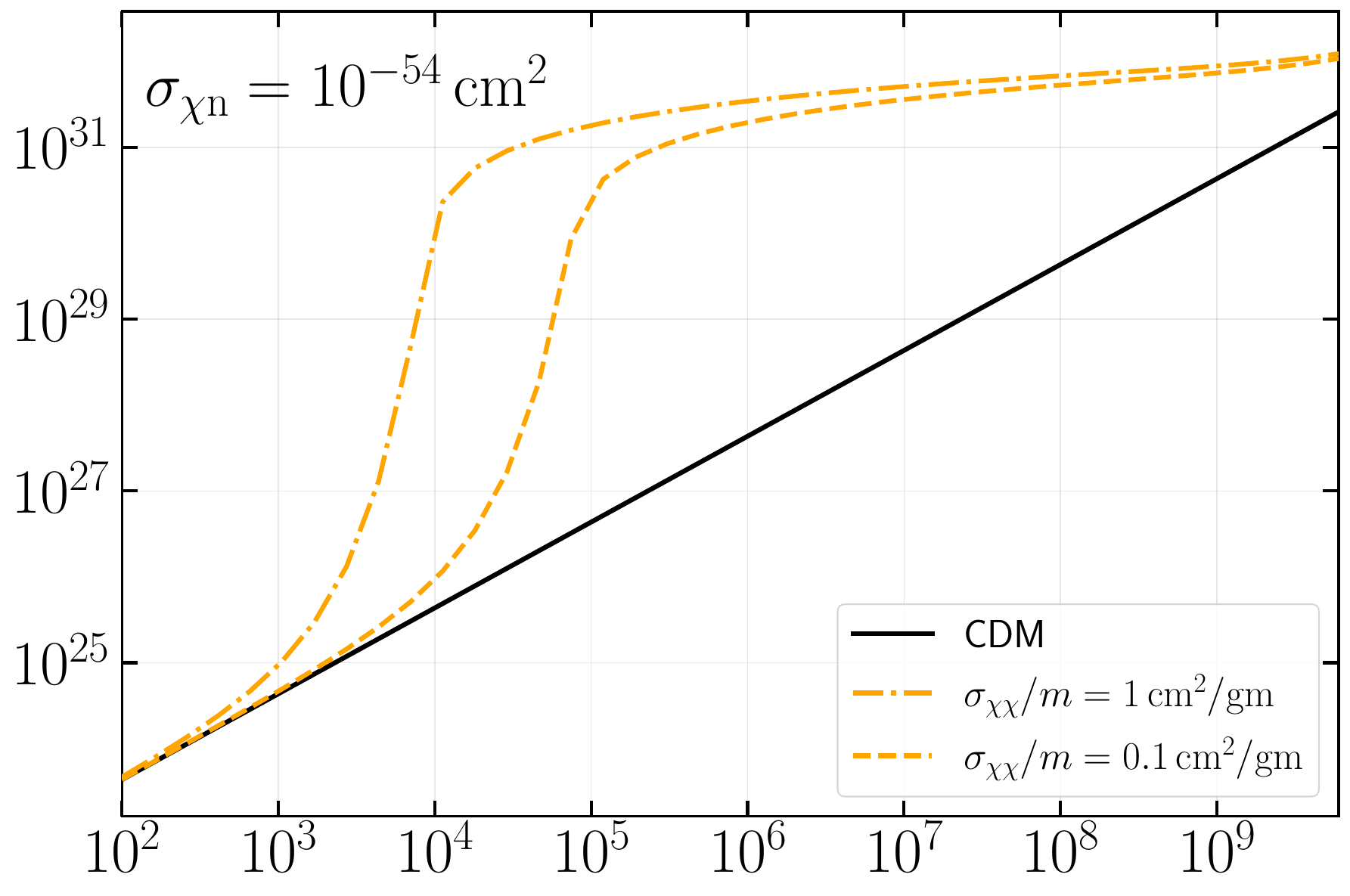}\\
		\includegraphics[scale=0.25]{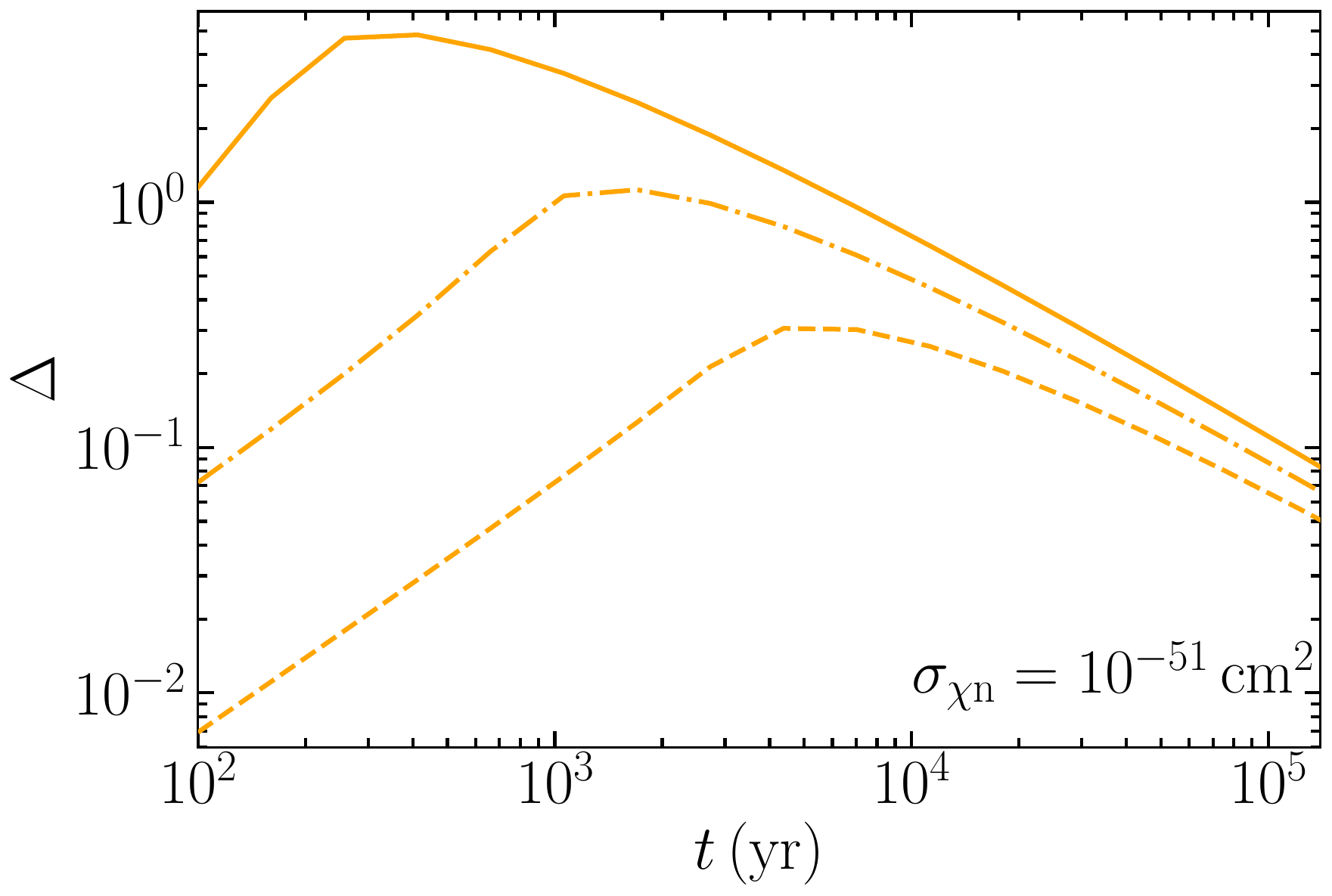}
		\includegraphics[scale=0.25]{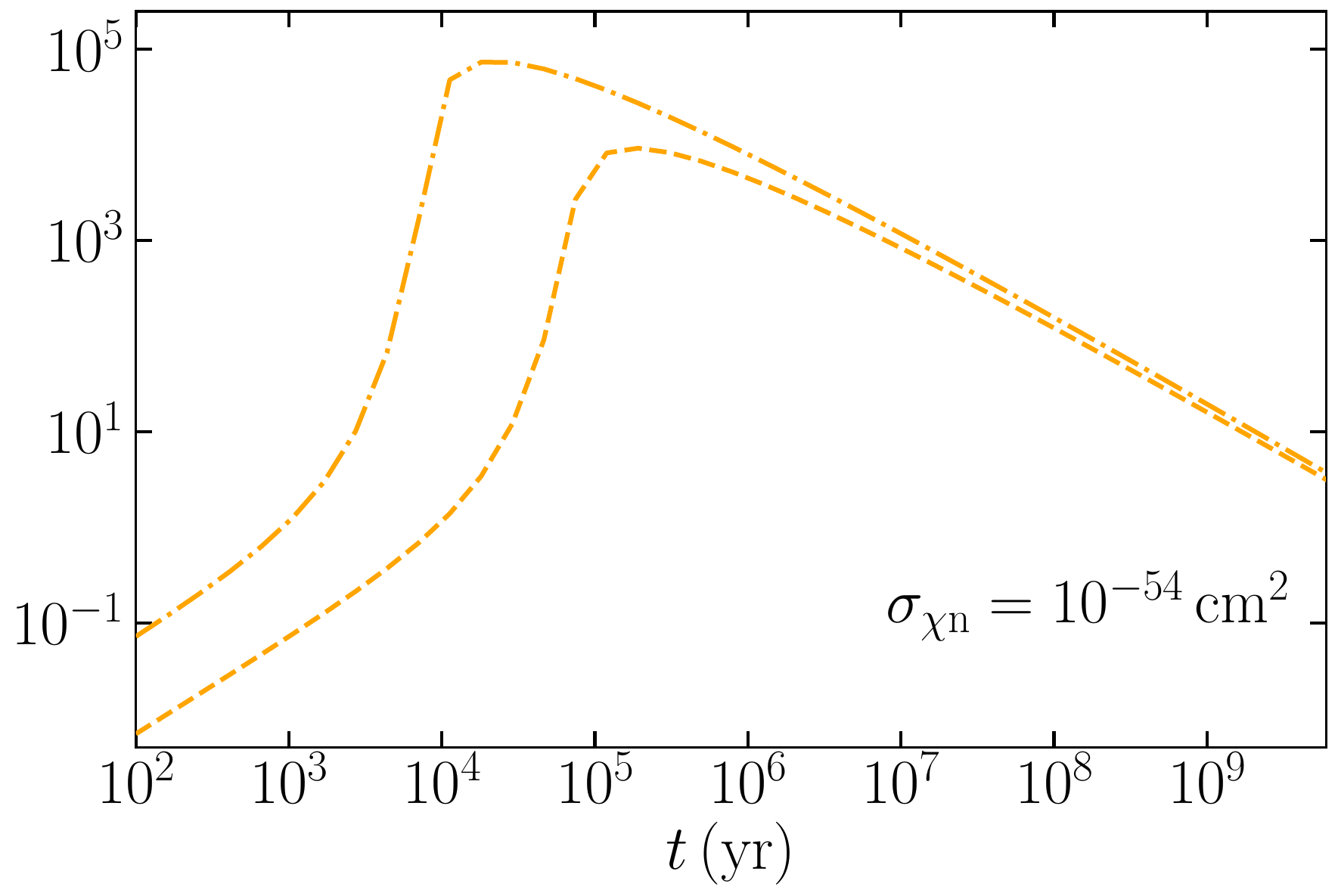}
	\end{center}
	\caption{\textit{Left}: Temporal variation in the captured DM particles (upper panel) and increment parameter (lower panel) for $\sigma_{\chi \rm n}=10^{-51}\rm\,cm^2$ and DM mass 100 GeV, in presence of DM self-interactions. The solid, dashed-dotted and dashed yellow curves represent $\sigma_{\chi \chi}/m =10\,,1.0$ and $0.1\rm\,cm^2/gm$ respectively. The black solid line is for $\sigma_{\chi \chi}/m =0$. \textit{Right}: Same as the left panel but for $\sigma_{\chi \rm n}=10^{-54}\rm\,cm^2$. Here we do not show the $\sigma_{\chi \chi}/m =10\, \rm cm^2/gm$ curve.}
	\label{fig:Numbert1}
\end{figure}
In this work we consider a typical NS of mass ($M_*$) $1.5 \, M_{\odot}$, radius ($R_*$) $12$ km and age ($t_{*}$) 1 Gyr \cite{Bell:2023ysh}. We show the evolution of $N_{\chi}(t)$ in figure \ref{fig:Numbert1} as a function of time, for certain benchmark values of $\sigma_{\chi \rm n}$ and $\sigma_{\chi \chi}$, and for a DM mass of $100$ GeV in the upper panel. In the lower panel we show the temporal variations in increment parameter $\Delta$, defined as,
\begin{equation}
\label{eq:delta}
\Delta(t)= \frac{N_{\chi}(\sigma_{\chi \chi} /m \neq 0,t)-N_{\chi}(\sigma_{\chi \chi} /m=0, t)}{N_{\chi}(\sigma_{\chi \chi} /m=0, t)}.
\end{equation}
Variation of $N_{\chi}(t)$ and $\Delta(t)$ for $\sigma_{\chi \chi} /m=1$ and $0.1\,\rm cm^2/gm$ are shown by the dashed-dotted and dashed curves respectively for $\sigma_{\chi \rm n}=10^{-51}$ (left panel) and $10^{-54}\rm \,cm^2$ (right panel). For $\sigma_{\chi \rm n}=10^{-51}\rm \,cm^2$, we show an additional solid yellow curve  with $\sigma_{\chi \chi} /m=10\,\rm cm^2/gm$.
Expectedly, we find an additional DM capture if self-interactions are present, and the increment in comparison to the CDM counterpart increases as we move below $\sigma_{\rm sat}$. This can be attributed to the fact that for low $\sigma_{\chi \rm n}$, $C_{\rm s}N_{\chi} \sim C_{\rm c}$, whereas for $\sigma_{\chi \rm n}$ near the saturation regime $C_{\rm s}N_{\chi} \ll C_{\rm c}$ \cite{Dasgupta:2020dik}. Therefore, in order to realize the impact of SIDM on NS temperatures, we work in the optically thin limit with cross-sections $<10^{-49}\rm \,cm^2$.

Given the enhanced number of DM that can accumulate inside a NS in presence of self-interactions, there remains a possibility for the captured DM to overcome the Chandrasekhar limit or form self-gravitating objects which can transmute the star into a black hole (BH) \cite{McDermott:2011jp,Guver:2012ba}. Additionally, for boson DM, there also exists the possibility of forming Bose-Einstein condensates (BEC) under low core temperatures \cite{McDermott:2011jp}. Admittedly, for the regions of DM-nucleon parameter space considered in this work it is not feasible to have BHs or BEC formation \cite{Kouvaris:2011fi,Saha:2025fgu}. Also as argued in \cite{Bell:2013xk} even the presence of extremely weak and repulsive self-interactions are enough to prevent BH formation.
\section{Impact of DM kinetic heating on neutron stars}
\label{sec:Kineticheat}

DM capture in NS has been extensively explored for analyzing and constraining particle properties of DM \cite{Bramante:2023djs}.
\begin{figure*}[t]
	\begin{center}
		\includegraphics[scale=0.3]{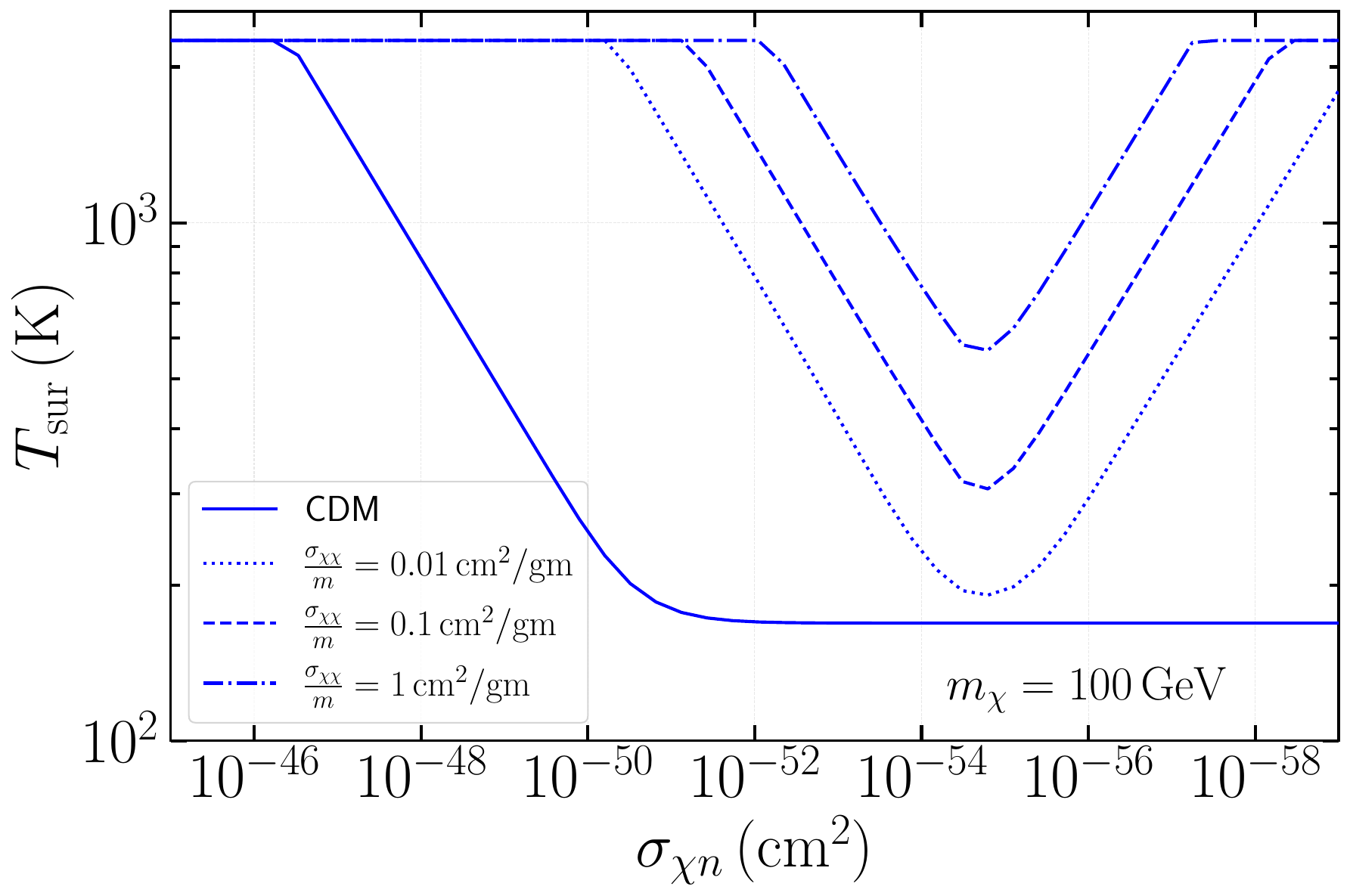}
		\caption{Variation of the surface temperatures with $\sigma_{\chi \rm n}$, for DM mass 100 GeV. The solid, dotted, dashed and dashed-dotted curves represent $\sigma_{\chi \chi}/m=0$ (CDM), $0.01,\,0.1$ and $1.0\,\rm cm^2/gm$ respectively.}
	\label{fig:chinvsm}
	\end{center}
\end{figure*}
Stars in the mass range of $(8 \sim 10\,) M_{\odot}$ undergoing supernovae explosions at the end of their nuclear fusion cycles, collapse into dense compact structures composed of degenerate neutrons. The core of a NS accounts for approximately $99\%$ of its mass with the inward gravitational pull being balanced by the neutron degeneracy pressure \cite{Sagert:2005fw}. It has been argued that DM deposits about $99\%$ of its kinetic energy onto the NS within a few hundred years \cite{Bell:2023ysh}. In this section we discuss the mechanism of heat deposition onto a NS by DM scattering with the nucleons and its subsequent effect on the NS cooling curve.
\subsection{Dark kinetic heating}
\label{subsec:Kineticheat}
Considering perfect black body (BB) radiation inside the NS, rate of DM kinetic energy injection over time is given by
\begin{equation}
\dot{E}_{\chi}^{\rm kin} = m_{\chi} \left( \frac{1}{\sqrt{B(0)}}-1 \right) C_{\rm geom} f,
\label{eq:energyrate}
\end{equation}
where $f=\text{min}[1, \sum_{i} \frac{C_i}{C_{\rm geom}}]$. The summation is performed over all species through which DM capture and thermalization occurs. The factor $f$ ensures that the capture rate does not exceed the saturation value with $C_{\rm geom}$ being the maximum capture rate at saturation. Temperature measured by an observer far from the star is given $T_{\chi}^{\rm obs}=\left(B^2(R_*)\,\dot{E}^{\rm kin}_{\chi} /4\pi R_*^2 \sigma_{B}\right)^{0.25}$, where symbols carry the usual meaning with $\sigma_{B}$ being the Stefan-Boltzmann constant. The values of $B$ at $r=0$ and $r=R_*$ are taken from \cite{Bell:2023ysh} for the quark-meson coupling (QMC) model \cite{Guichon:2018uew}. We now compute the effect of dark kinetic heating for asymmetric DM on the NS cooling curves.
\subsection{Implications on neutron star cooling}
\label{subsec:nscooling}
\begin{figure}[t]
	\begin{center}
		\subfloat[\label{sf:Ttg3}]{\includegraphics[scale=0.26]{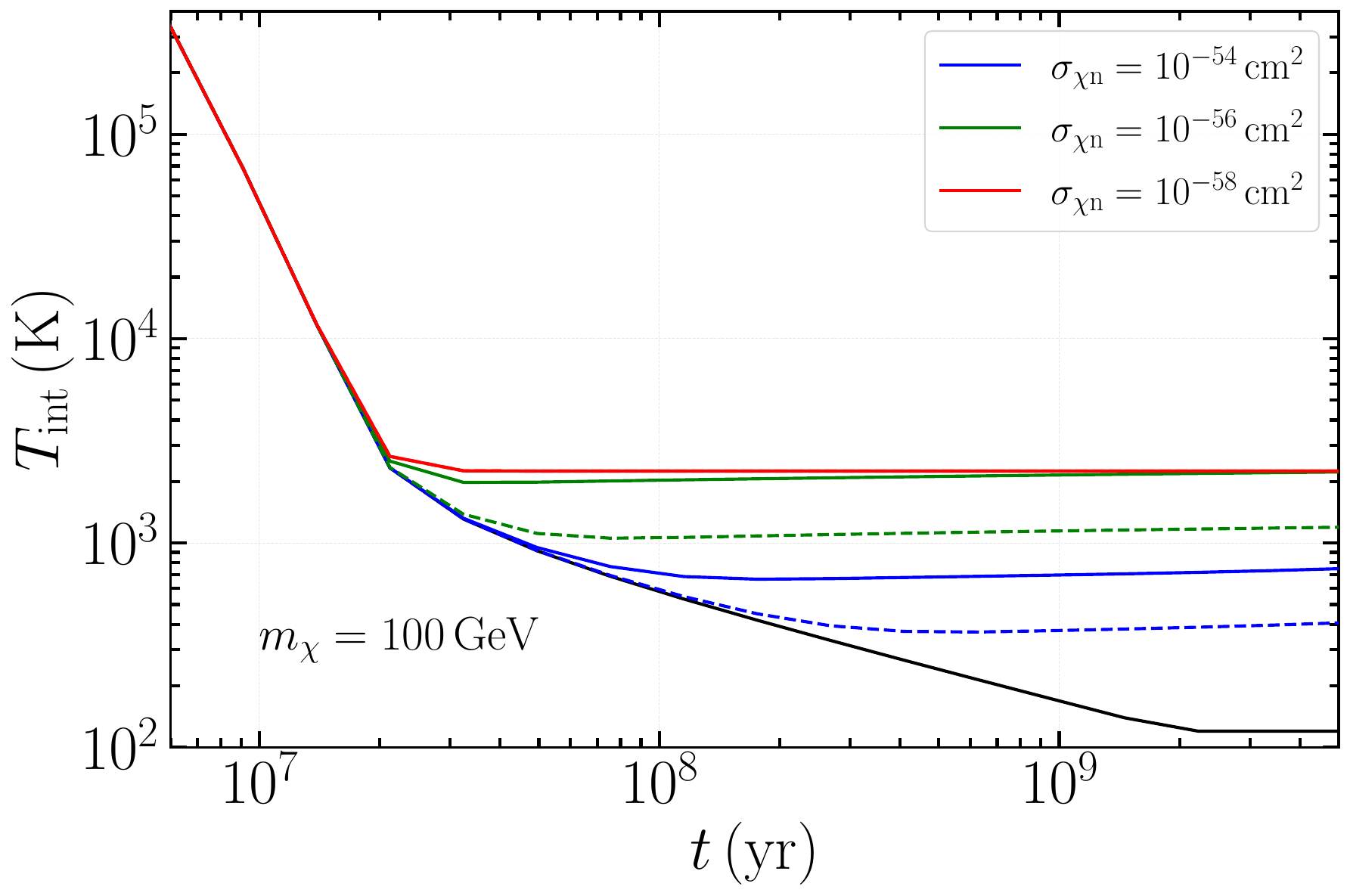}}
		\subfloat[\label{sf:Ttg4}]{\includegraphics[scale=0.26]{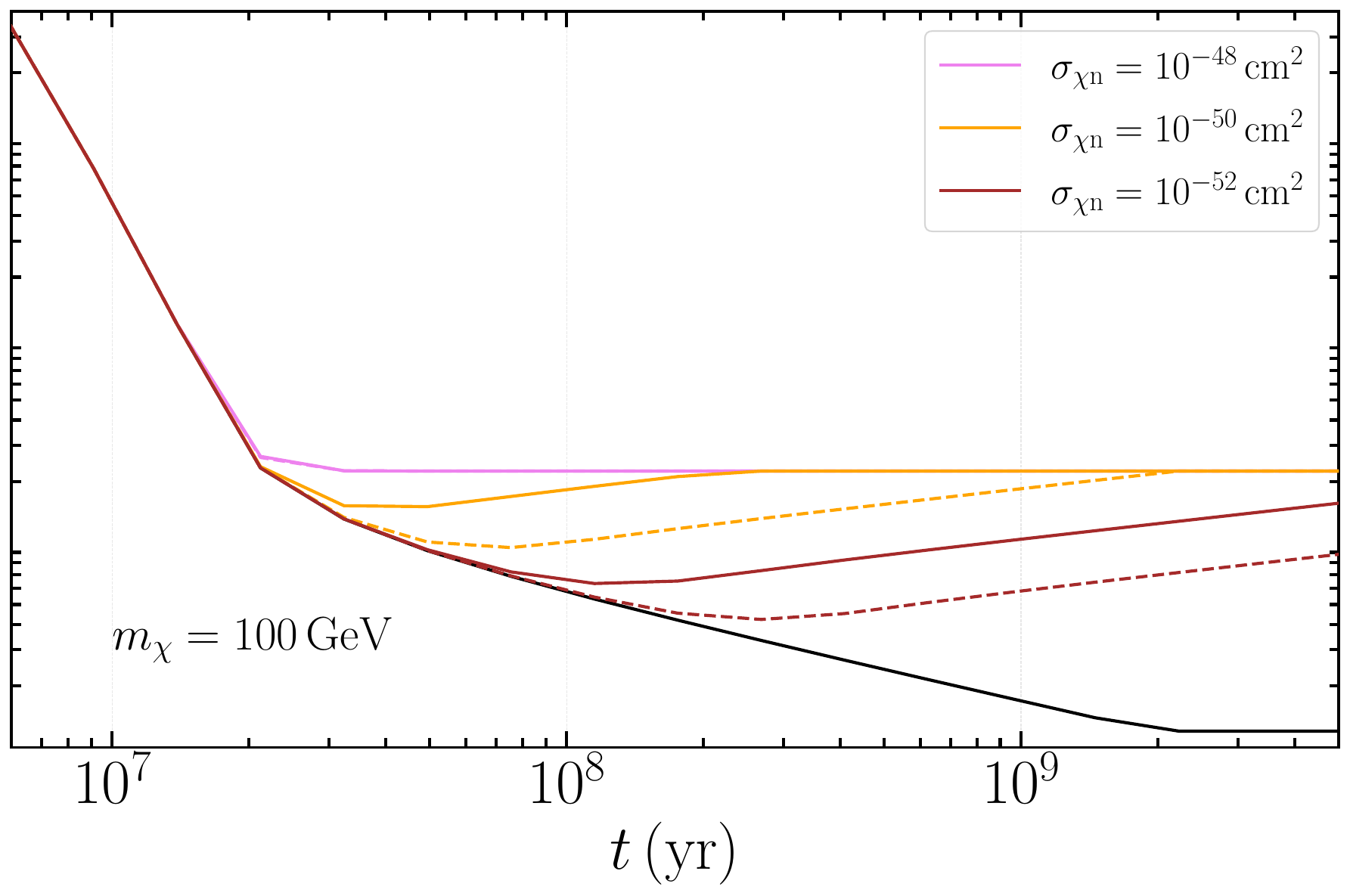}}
	\end{center}
	\caption{Evolution of $T_{\rm int}(t)$ of a NS in the presence of DM self-interactions for a DM mass of 100 GeV. DM-nucleon $\sigma_{\chi \rm n}=10^{-54}\,,10^{-56}$ and $10^{-58}\rm \,cm^2$ are given by the blue, green and red curves respectively in the left panel \ref{sf:Ttg3}, for $t_{\rm th}>t_{*}$. The right panel \ref{sf:Ttg4}, is for $t_{\rm th}<t_{*}$, where $\sigma_{\chi \rm n}=10^{-48}\,,10^{-50}$ and $10^{-52}\rm \,cm^2$ given by the violet, orange and brown curves respectively. The solid black curve represents the standard NS cooling scenario without any self-interaction. The solid and dashed colored curves in both plots are representative of $\sigma_{\chi \chi}/m=1$ and $0.1\,\rm cm^2/gm$ respectively.}
	\label{fig:Tt}
\end{figure}
After a NS is formed from supernovae explosions, it undergoes a phase of thermalization and cooling, mediated by the emission of neutrinos and photons \cite{1983bhwd.book.....S,Yakovlev:1999sk,Yakovlev:2004iq,Page:2004fy,Ofengeim:2017cum,Potekhin:2020ttj}.  It takes a young NS roughly $100$ yrs to relax and thermalize \cite{1994ApJ...425..802L,Gnedin:2000me,Fortin:2009ry}. We therefore initialize the internal temperature as $T_{\rm int}(t=100\, \text{yrs})=10^9$ K. Thermal evolution of the NS temperature $T_{\rm int}$, is described by the differential equation 
\begin{equation}
\frac{dT_{\rm int}}{dt}=\frac{\epsilon_\chi - \epsilon_\nu -\epsilon_\gamma}{c_{\rm V}},
\label{eq:Tt}
\end{equation}
where $\epsilon_\chi$, $\epsilon_\nu$ and $\epsilon_\gamma$ are the emissivity due to DM, neutrino and photon respectively and $c_{\rm V}$ is the NS heat capacity. At times $\leq10^5$ yrs, neutrino emission from core dominates the cooling mechanism through direct Urca \cite{1941PhRv...59..539G}, modified Urca \cite{1995A&A...297..717Y} and Pair-Breaking and Formation (PBF) \cite{Senatorov:1987aa} processes, whereas the later phase is associated with cooling via photon emission. In solving equation \eqref{eq:Tt} we take the relevant expressions from \cite{1983bhwd.book.....S,Kouvaris:2007ay}. In this work we consider DM kinetic heating from asymmetric DM as the major source of energy injection, neglecting energy release from annihilation. The dark emissivity using equation \eqref{eq:energyrate} is given by $\epsilon_\chi= 3\,\dot{E}_{\chi}^{\rm kin}/4 \pi R_*^3$ \cite{Baryakhtar:2017dbj}. $T_{\rm int}$ hence obtained can be translated to the NS surface temperature $T_{\rm sur}$. However for temperatures less than $3700$ K distinction between the two is insignificant. \cite{Page:2004fy,1982ApJ...259L..19G}. In figure \ref{fig:chinvsm} we show the variation of $T_{\rm sur}$,
with the DM-nucleon interaction cross-section, for four values of specific self-interaction cross-section including CDM. We note that as $\sigma_{\chi \rm n}$ decreases, initial consistency in the NS surface temperature is maintained by DM self-interactions, finally before dropping with $\sigma_{\chi \rm n}$. However, as $\sigma_{\chi \rm n}$ becomes less than a certain cross-section $\sigma_{\rm t}$ (defined here as the turn-over cross-section), $T_{\rm sur}$ starts to increase due to DM self-interactions. As $\sigma_{\chi \rm n}$ decreases, the thermalization time starts to increases and can go well beyond the age of NS. For values of $\sigma_{\chi \rm n} < \sigma_{\rm t}$, DM orbits given by $r_{\chi}(t)$ do not reach the core of NS within $t_*$, as $t_{\rm th}> t_{*}$. These non-thermalized DM orbiting at radius larger than $r_{\rm th}$ increase the extent of the DM distribution inside the NS core, relative to the thermalization radius. For such low values of $\sigma_{\chi \rm n}$, the DM which subsequently enter the star do not require to migrate towards the NS core, and can find other DM particles to scatter with and can deposit energy at radius well beyond $r_{\rm th}$ \cite{Chen:2018ohx}. In figure \ref{fig:chinvsm}, it is due to this effect of self-interactions between the non-thermalized DM, we see an increase in the captured DM particles and hence $T_{\rm sur}$, even as $\sigma_{\chi \rm n}$ decreases beyond $\sigma_{\rm t}$.

In figure \ref{fig:Tt} we show the temporal evolution of the internal NS temperature $T_{\rm int}$ for 100 GeV DM mass, for certain benchmark values of $\sigma_{\chi \rm n}$ and $\sigma_{\chi \chi}$. The solid black curve represent the standard NS cooling scenario. The blue, green, and red curves in the left panel are for $\sigma_{\chi \rm n}=10^{-54}\rm \,cm^2$, $\sigma_{\chi \rm n}=10^{-56}\rm \,cm^2$ and $\sigma_{\chi \rm n}=10^{-58}\rm \,cm^2$ respectively, all of which have thermalization times greater than the age of NS. Whereas in the right panel, the violet, orange, and brown curves are for $\sigma_{\chi \rm n}=10^{-48}\rm \,cm^2$, $\sigma_{\chi \rm n}=10^{-50}\rm \,cm^2$ and $\sigma_{\chi \rm n}=10^{-52}\rm \,cm^2$ respectively, for which thermalization times are less than the age of NS. The solid and dashed curves for each $\sigma_{\chi \rm n}$ are plotted for $\sigma_{\chi \chi}/m=1$ and $0.1 \,\rm cm^2/gm$ respectively. In figure \ref{sf:Ttg3}, self-interaction between the non-thermalized DM orbiting at larger radii, compared to the thermalized DM, cause a significant increase in the DM induced kinetic heating and hence the internal temperature $T_{\rm int}$. This can be understood by drawing parallels with figure \ref{fig:chinvsm}. The DM orbital radius being inversely related to $\sigma_{\chi \rm n}$ \cite{Guver:2012ba}, $r_{\chi}$ and hence $T_{\rm int}$ increase with decreasing $\sigma_{\chi \rm n}$, for non-thermalized DM. Overall figure \ref{fig:Tt} shows that DM self-interactions can cause appreciable heating in old neutron stars to surface temperatures $\mathcal{O}(1000)$ K, absent for collision-less DM. Admittedly for specific NS models, such temperatures may also arise from vortex creeping effects, making it difficult to identify the heat source \cite{Fujiwara:2023hlj}. The question now remains whether such temperatures can be probed by present or future infrared telescopes.
\section{Prospects for NS detection and SIDM constraint}
\label{sec:detec}

In this section we explore the prospect of identifying an old NS of age $t_*$ and $T_{\rm sur}\sim 1000$ K. Thereby constraining the DM parameter space ($\sigma_{\chi \rm n},\ \sigma_{\chi \chi}$ and $m$), essential for dark kinetic heating in asymmetric DM. Neutron stars observed in the UV regime are ideal for probing DM annihilation effects \cite{deLavallaz:2010wp}. However, infrared telescope like the James Webb Space Telescope (JWST) can potentially look into low heating signatures arising solely from DM scattering \cite{Baryakhtar:2017dbj}. It was pointed out in \cite{Baryakhtar:2017dbj} that old and isolated neutron stars of $\mathcal{O}(1000)$ K are excellent targets for detecting DM heating signatures. Monte-Carlo orbital simulations of galactic neutron stars predict 1-2 (100–200) cold, old and isolated neutron stars within a distance of 10 (50) pc \cite{2009PASP..121..814O,2010A&A...510A..23S}. However, it was recently shown in \cite{Bramante:2024ikc,Raj:2024kjq} that there are no detectable neutron stars present within a distance of 10 pc from Earth and with temperatures of 2000 K.The currently operating JWST \cite{Gardner:2006ky}, along-with proposed European Extremely Large Telescope (ELT) \cite{2018sf2a.conf....3N} and Thirty Meter Telescope (TMT) \cite{TMTInternational:2015pvw} are capable of detecting BB temperatures of 1300–4300 K with their optical to near-infrared imaging instruments \cite{Das:2024thc}. The PSR J2144-3933 which is reported to be one of the coldest NS observed, with an approximate spin-down age of $3 \times 10^8$ yrs is a slow, isolated and rotation powered pulsar located approximately 170 pc from Earth \cite{Page:2005fq,Ofengeim:2017cum}. Its effective temperature reported by the Hubble Space Telescope is 33000 K \cite{Guillot:2019ugf}, which can be further constrained to 20000 K with JWST-NIRCam, 15000 K with ELT-MICADO, and 9000 K with TMT-IRIS \cite{Raj:2024kjq}. Other cold and old neutron stars located in the solar neighborhood include PSR J0437-4715 \cite{McDermott:2011jp} and PSR J0108-1431 \cite{Abramkin:2021tha}, roughly 130 pc from Earth. Even with these cold neutron stars and advanced telescopes, measurement of low dark heating temperatures $\sim 1000$ K remains a challenge.

Admittedly, probing such SIDM parameters would require DM-neutron cross-sections well into the neutrino fog. DM experiments such as XENONnT, PandaX and LUX-ZEPLIN (LZ) set the leading constraints on $\sigma_{\chi \rm n}$ with nuclear recoil, for both light \cite{XENON:2024hup,PandaX:2025rrz,LZ:2025igz} and heavy\cite{LZ:2022lsv,XENON:2025vwd} WIMP-like DM at $\sigma_{\chi \rm n} < 10^{-45} \rm cm^2$ for $m_{\chi}\sim5$ GeV and $\sigma_{\chi \rm n} < 10^{-47} \rm cm^2$ for $m_{\chi}\sim 30$ GeV respectively. The TESSERACT \cite{TESSERACT:2025tfw} and CRESST \cite{CRESST:2024cpr} collaborations report some of the strongest bounds on $\sigma_{\chi \rm n}$ for sub-GeV DM. Experiments such as SuperCDMS SNOLAB \cite{SuperCDMS:2022kse} and DARWIN \cite{DARWIN:2016hyl} are expected to reach the neutrino fog in near future. Sensitivity to such low cross-sections are limited by neutrino backgrounds, where statistical discrimination between the two become increasingly difficult. Further progress would require new detection strategies where solar neutrino, atmospheric neutrino and Diffuse Supernova Neutrino Background (DSNB) fluxes act as irreducible backgrounds \cite{DeRomeri:2025nkx}.

We now discuss the possibility of constraining DM self-interaction from kinetic heating, based on the potential detection of a NS with observed surface temperatures in the range 1000-1200 K. We present our results in two regimes, demarcated by the thermalization timescale.
\begin{enumerate}
\item{\textbf{Thermalization time lower than age of neutron star:}}
\label{subsec:detec1}
For DM-nucleon cross-sections below $\sigma_{\rm sat}$ and larger than $\sigma_{\rm t}$, thermalization occurs within the NS age.
\begin{figure*}[t]
	\begin{center}
		\subfloat[\label{sf:C1}]{\includegraphics[scale=0.28]{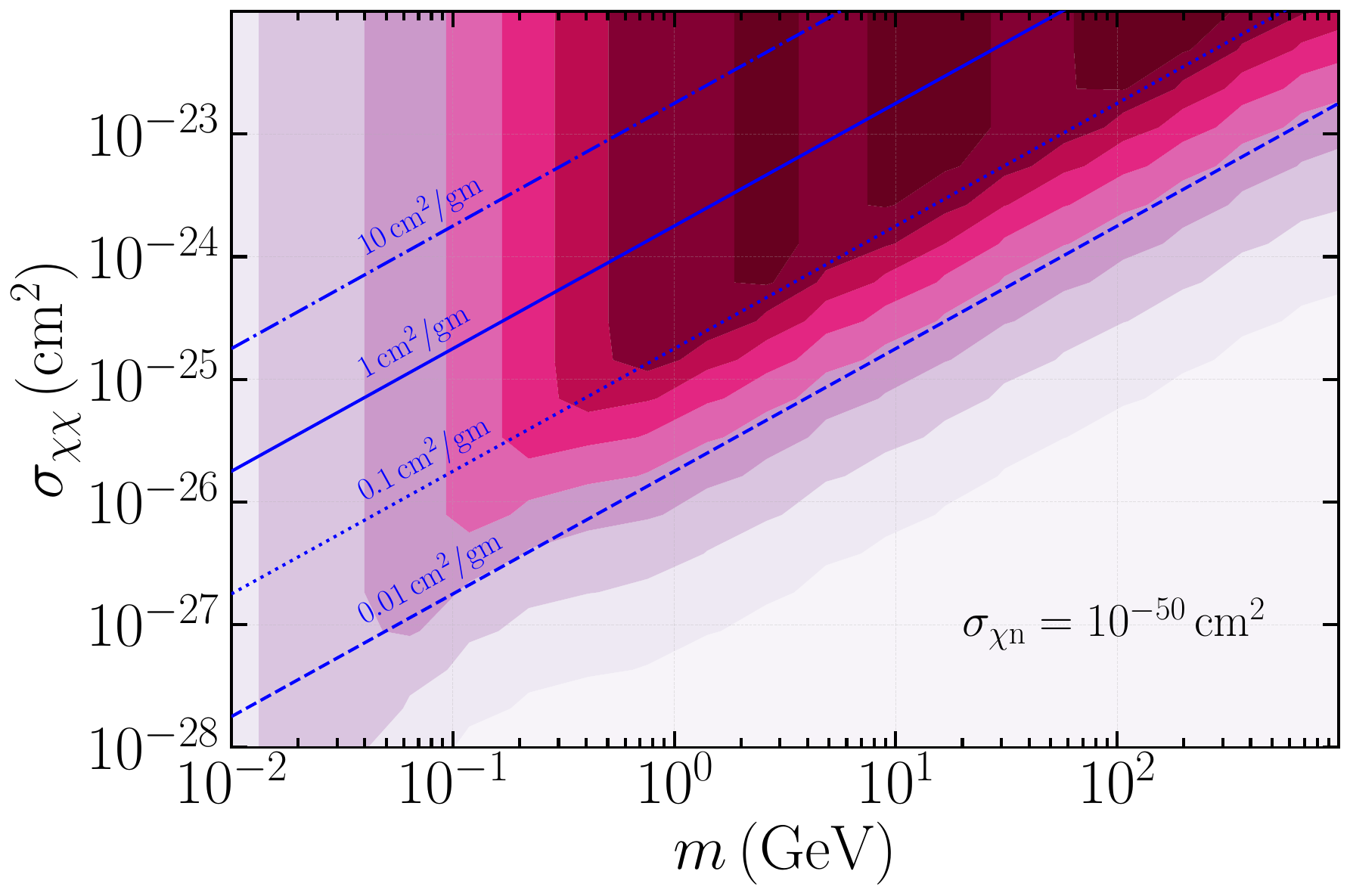}}
		\subfloat[\label{sf:C2}]{\includegraphics[scale=0.28]{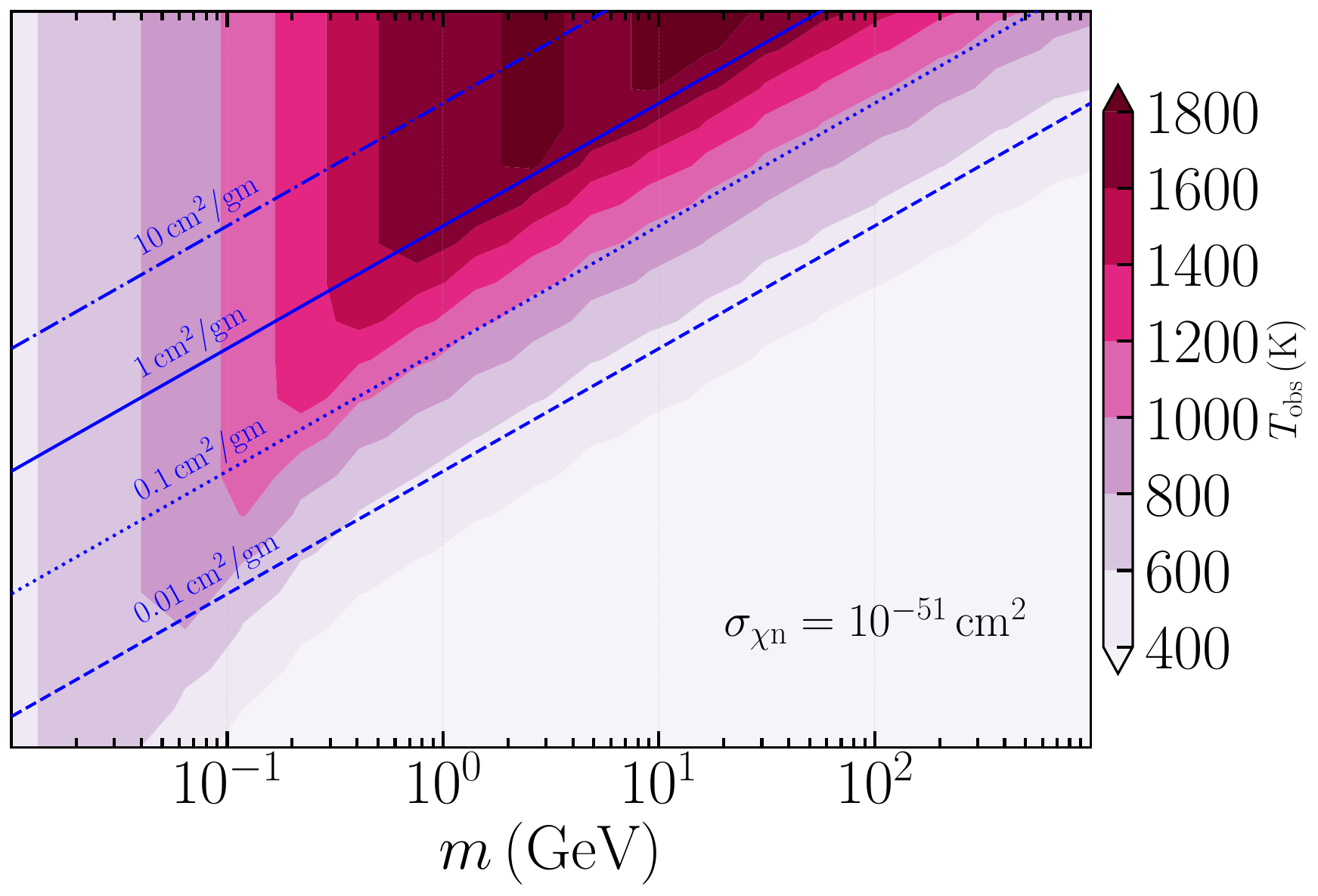}}
		\caption{Contours of observed surface temperatures (in Kelvin), scanned over self-interaction cross-section $\sigma_{\chi \chi}$ and DM mass, for $\sigma_{\chi \rm n}=10^{-50}\, \rm cm^2$ (\ref{sf:C1}), and $10^{-51}\, \rm cm^2$ (\ref{sf:C2}). The blue curves represent the contours of specific self-interaction cross-section, $\sigma_{\chi \chi}/m=0.01,\,0.1,\,1$ and $10\,\rm cm^2/gm$.}
		\label{fig:contour1}
	\end{center}
\end{figure*}
In figure \ref{fig:contour1} we plot the observed surface temperature contours $T_{\rm obs}$ for $\sigma_{\chi \chi}$ vs. DM mass with $\sigma_{\chi \rm n}=10^{-50}\rm \,cm^2$ and $10^{-51}\rm \,cm^2$ in the left and right panels respectively, where $T_{\rm obs}=\sqrt{B(R_*)}\,T_{\rm sur}$ as discussed in section \ref{subsec:Kineticheat}. We report that even for low values of $\sigma_{\chi \rm n} < 10^{-49}\rm \,cm^2$, certain combinations of DM mass and self-interaction cross-section can give rise to $T_{\rm obs}$ between 1000-1200 K. The potential detection of an old NS with these temperatures can set a lower bound on the specific SIDM cross-section to $0.01\,(0.1)\,\rm cm^2/gm$, for $\sigma_{\chi \rm n}=10^{-50}\rm \,cm^2\,(10^{-51}\rm \,cm^2$) and for DM in the mass range 100 MeV to 1 TeV. Therefore cold neutron stars can provide a lower bound on the specific SIDM cross-section two orders of magnitude stringent than the bullet cluster. Evidently, the bounds on $\sigma_{\chi \chi}/m$ that we arrive at are sensitive to the DM-neutron cross-section.
\item{\textbf{Thermalization time greater than age of neutron star:}}
\label{subsec:detec2}
\begin{figure*}[t]
	\begin{center}
		\subfloat[\label{sf:C3}]{\includegraphics[scale=0.28]{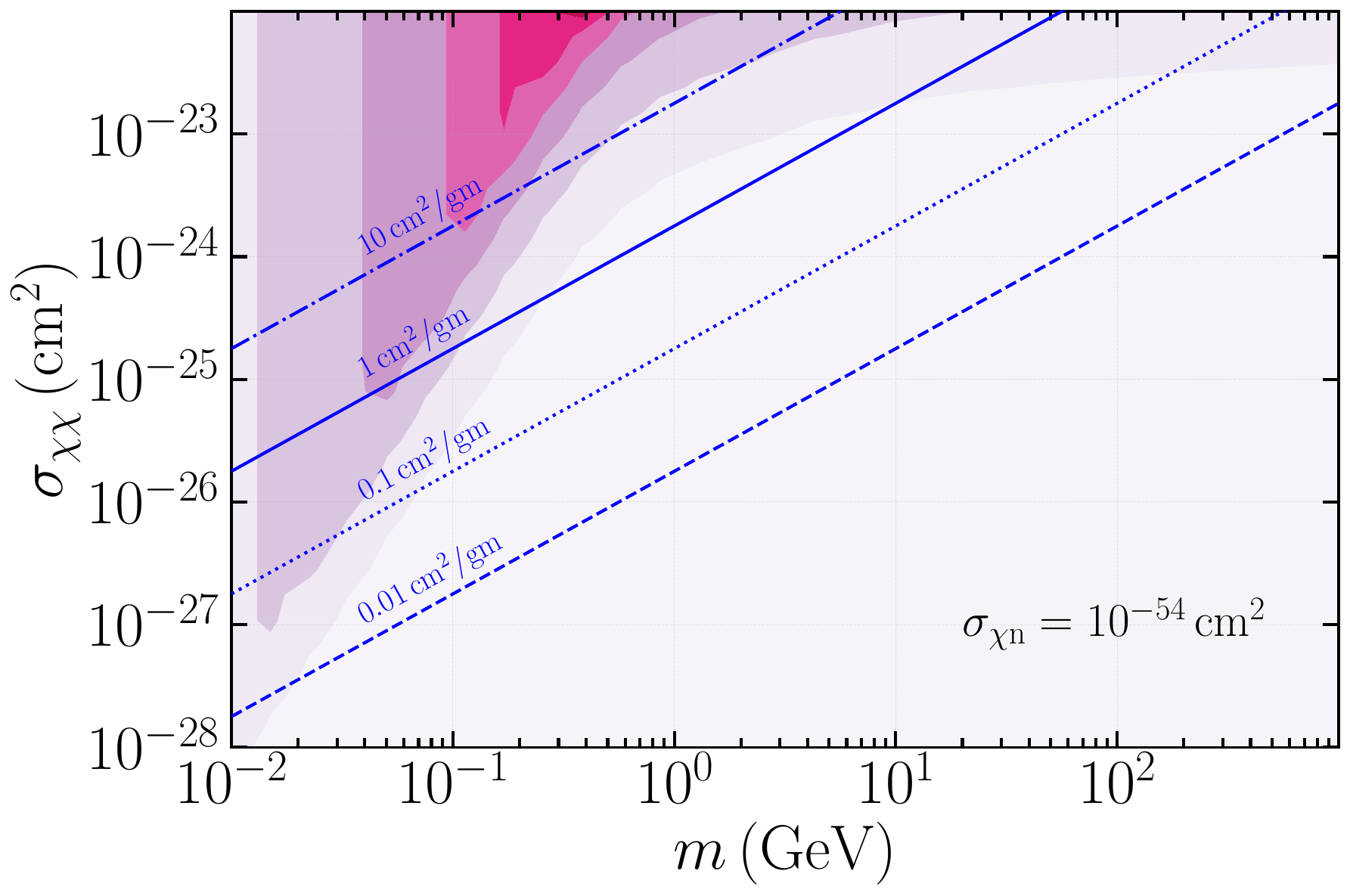}}
		\subfloat[\label{sf:C4}]{\includegraphics[scale=0.28]{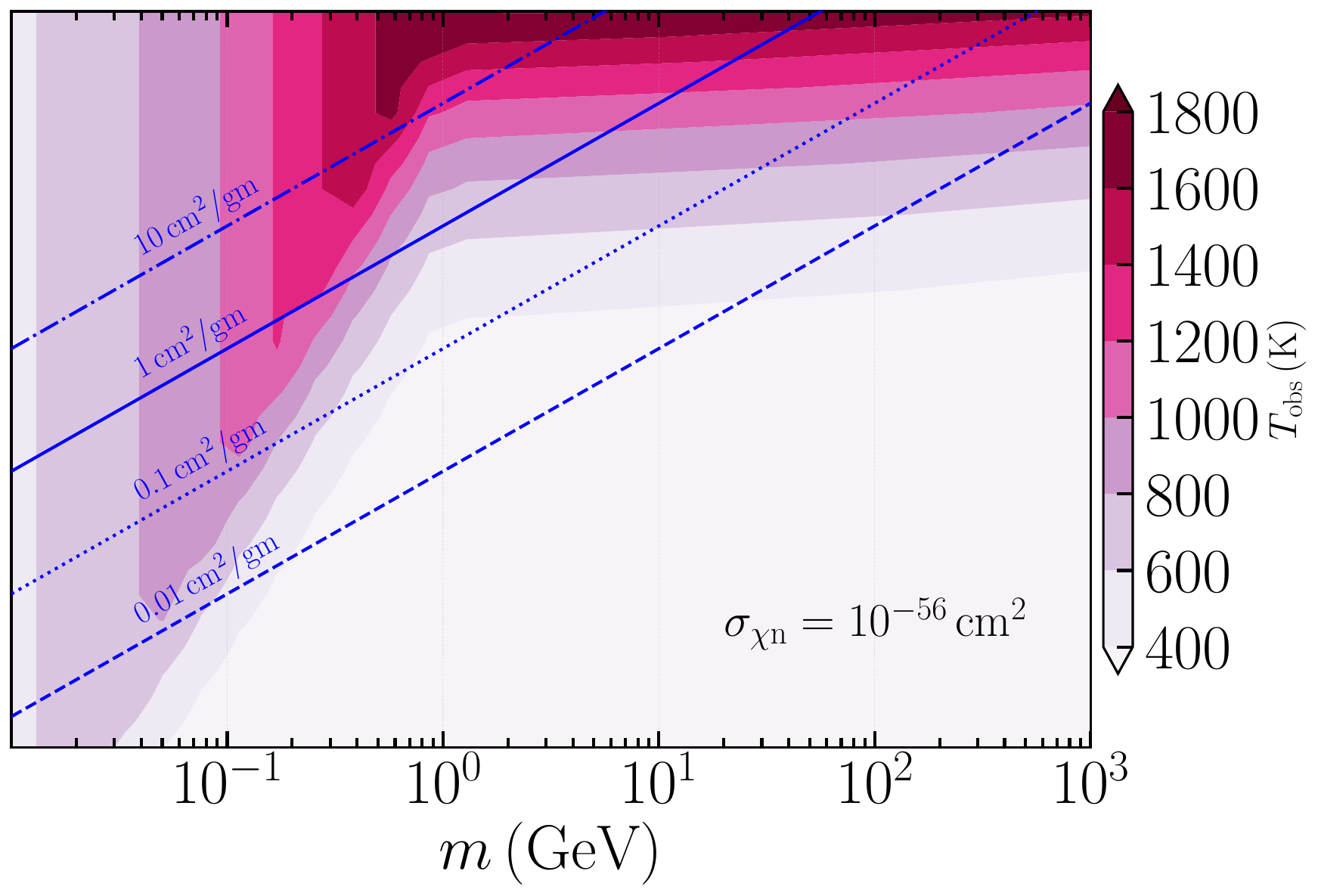}}
		\caption{Similar to figure \ref{fig:contour1}. The left (\ref{sf:C3}) and right (\ref{sf:C4}) panels are plotted for $\sigma_{\chi \rm n}=10^{-54}\, \rm cm^2$ and $10^{-56}\, \rm cm^2$ respectively with $t_{\rm th}>t_*$.}
		\label{fig:contour2}
	\end{center}
\end{figure*}
We now turn our attention to $\sigma_{\chi \rm n}$ for which thermalization timescales are larger than $t_*$. In figure \ref{fig:contour2}, we plot contours of SIDM cross-section vs. DM mass for two values of $\sigma_{\chi \rm n}$ namely $10^{-54}\rm \,cm^2$ and $10^{-56}\rm \,cm^2$ in the left and right panels respectively. Here we note an increase in $T_{\rm obs}$ for $\sigma_{\chi \rm n}=10^{-56}\, \rm cm^2$ in comparison to $10^{-54}\rm \,cm^2$. As seen in figure \ref{fig:chinvsm}, for $\sigma_{\chi \rm n} < \sigma_{\rm t}$, the surface temperatures begin to rise aided by self-interactions. These however are DM existing at radii much larger than the NS core radius \cite{Guver:2012ba}. Consequently, the DM hence captured may deposit its energy without migrating to the NS core \cite{Chen:2018ohx}, leading to an increased surface temperature at such low DM-neutron cross-sections. However, such low $\sigma_{\chi \rm n}$ are beyond the reach of present or proposed direct detection experiments. 

For low values of $\sigma_{\chi \rm n}$, capture of DM by neutrons reduce drastically. The capture rate further reduces for massive DM owing to its low number density. In this regime DM self-interactions become significant $C_{\rm s}N_{\chi}\sim C_{\rm c}$, and is solely responsible for rising $T_{\rm obs}$ to $1000$ K. Although it would be difficult to provide constraints on SIDM in this regime, it might be interesting  to probe lower values of the DM parameter space well inside the neutrino fog, utilizing low values of $\sigma/m$ in neutron stars with temperatures $\mathcal{O}(1000)$ K.
\end{enumerate}

In deriving these bounds, we remained agnostic regarding the particle physics model. However, it is worth mentioning that interactions of fermionic DM, mediated by light scalar or vector particles \cite{Buckley:2009in,Tulin:2013teo,Zurek:2013wia,Patel:2022qvv} can account for $\sigma_{\chi \chi}$ in this range. Apart from these, bosonic \cite{Bramante:2013hn,Karkevandi:2021ygv,Giangrandi:2022wht} and fermionic \cite{Kouvaris:2011gb,Modak:2015npa,Borah:2021pet} DM models have been extensively explored to account for DM self-interactions. Although we consider velocity-independent SIDM cross-sections, the constraints we arrive at lie in the ballpark of existing bounds derived for velocity dependent DM self-interaction cross-sections from the bullet cluster \cite{Randall:2008ppe}, stellar dispersion and lensing measurements of cores in clusters \cite{2018ApJ...853..109E}, subhalo mergers \cite{Harvey:2015hha} and shape of DM halos from gravitational lensing of clusters \cite{Peter:2012jh}, all with relative DM velocities $\mathcal{O}(1000)\, \rm km/s$. Given the semi-relativistic velocities with which DM fall onto neutron stars $\sim 0.6$ c, our bounds show consistency with those derived from DM relative velocities, characteristic of galaxy cluster scales \cite{Adhikari:2022sbh}.
\section{Summary and conclusion}
\label{sec:conclude}

Dense compact objects like neutron stars have proven to be effective probes of low DM masses and weak interaction cross-sections. The discovery of faint neutron stars with surface temperatures $\sim 1000$ K may serve as conclusive evidence for NS heating by particle DM. Recent times witnessed a significant number of observations from JWST hinting at an improved statistics, which is likely to increase in upcoming infrared telescopes like ELT, TMT, as well as earth based observatories.

In this work we systematically consider the accumulation of self-interacting asymmetric DM inside a NS. Their capture and thermalization is associated with the transfer of kinetic energy to the constituents of the NS. These processes heat up the NS over its standard model predicted surface temperatures, particularly during late times $\sim 1$ Gyr. We find a significant enhancement in the accumulated SIDM particles relative to the CDM paradigm. The optically thin limit is particularly interesting in this regard. For DM-nucleon cross-sections in the range $\sigma_{\chi \rm n}<\sigma_{\rm sat}$, DM self-interactions aid in maintaining temperatures $\mathcal{O}(1000)$ K. We report that detection of an old NS with surface temperatures 1000-1200 K in present or upcoming telescopes will set a lower bound on the specific DM self-interaction cross-section $\sigma/m \geq 0.01\,(0.1) \, \rm cm^2/gm$, which is two (one) orders of magnitude stronger than the bullet cluster, given $\sigma_{\chi \rm n}\sim 10^{-50}(10^{-51})\,\rm cm^2$ and for DM in the mass range of 100 MeV and 1 TeV. For DM-nucleon cross-sections $\sigma_{\chi \rm \rm n}< \sigma_{\rm t}$, self-interactions between non-thermalized DM may effectively increase the NS surface temperatures to $\mathcal{O}(1000)$ K. The intriguing possibility of self-interactions between non-thermalized DM can probe cross-sections currently inside the neutrino fog, beyond the reach of direct DM searches. The detection of such cold neutron stars would serve as a smoking-gun signature for DM self-interactions with $\sigma_{\chi \rm n}$ in the specified limit. This however would depend on an accurate modeling of NS heating mechanism. 

A more quantitative estimate of the numbers provided in this work would arise from a detailed understanding of the NS cooling mechanism, which depends on vortex creeping effects, spin and magnetic properties of the NS and modeling of the NS core. In this work we explore the properties of NS as an astrophysical particle collider to present competitive bounds on DM self-interaction cross-section, with possible extensions of DM parameter space to the neutrino fog. It would be interesting to look into the physics of SIDM in neutron stars and white dwarfs, arising for various DM candidates e.g axions, sterile neutrinos, and dark photons. Possible extensions of this study can be made to multi-component DM which can subsequently be constrained using multi-messenger astrophysics.
\paragraph*{Acknowledgments:}

The author would like to thank Tirtha Sankar Ray, Ujjal Kumar Dey, Tarak Nath Maity, Debajit Bose, Nirmal Raj, Chang Sub Shin and Seodong Shin for the helpful discussions. This work is partially supported by the National Research Foundation of Korea (NRF) grant funded by the Ministry of Science and ICT (RS-2025-00562917).
\appendix
\bibliographystyle{JHEP}
\bibliography{nsheat.bib}

\end{document}